\documentclass[12pt]{article}

\usepackage{hyperref,subfig,booktabs,amsthm,
  natbib, amssymb, graphicx, amsmath,color, lineno}

\hypersetup{backref, colorlinks=false, filecolor=white}

\usepackage{Sweave}
\begin{document} 
\title{Employer Expectations, Peer Effects and Productivity:
  \\ Evidence from a Series of Field Experiments} \date{August 3, 2010}

\author{John J. Horton \\ Harvard University\footnote{ Email:
    john.joseph.horton@gmail.com.  Thanks to the NSF-IGERT
    Multidisciplinary Program in Inequality \& Social Policy (Grant
    No. 0333403), the University of Notre Dame and the John Templeton
    Foundation's Science of Generosity Initiative for generous
    financial support. Thanks to the Centre for Economic Performance
    at the London School of Economics and Political Science for
    hosting me while I worked on this project. Thanks to Richard
    Zeckhauser, Robin Yerkes Horton, Larry Katz, Nicholas Christakis,
    Rob Miller, Malcolm-Wiley Floyd, Renata Lemos and Lydia Chilton
    for helpful comments and suggestions. Thanks to John Comeau and
    Alex Breinen for research assistance. All datasets, code and
    auxiliary regression results are currently or will be available at
    the author's website:
    \href{http://sites.google.com/site/johnjosephhorton/}{http://sites.google.com/site/johnjosephhorton/}.}}

\maketitle

\begin{abstract} 
This paper reports the results of a series of field experiments
designed to investigate how peer effects operate in a real work
setting. Workers were hired from an online labor market to perform an
image-labeling task and, in some cases, to evaluate the work product
of other workers. These evaluations had financial consequences for
both the evaluating worker and the evaluated worker. The experiments
showed that on average, evaluating high-output work raised an
evaluator's subsequent productivity, with larger effects for
evaluators that are themselves highly productive. The content of the
subject evaluations themselves suggest one mechanism for peer effects:
workers readily punished other workers whose work product exhibited
low output/effort. However, non-compliance with employer expectations
did not, by itself, trigger punishment: workers would not punish
non-complying workers so long as the evaluated worker still exhibited
high effort. A worker's willingness to punish was strongly correlated
with their own productivity, yet this relationship was not the result
of innate differences---productivity-reducing manipulations also
resulted in reduced punishment. Peer effects proved hard to stamp out:
although most workers complied with clearly communicated maximum
expectations for output, some workers still raised their production
beyond the output ceiling after evaluating highly productive yet
non-complying work products.
\newline

JEL J01, J24, J3
\end{abstract}


\newcommand{\imgW}{.65}
\newcommand{\imgWs}{.3}
\setkeys{Gin}{width=\imgW \textwidth}

\section{Introduction} 
A perennial question of interest to both economists and firm managers
alike is why employees work hard despite facing weak incentives and
light monitoring. Many theories have been proposed: firms might obtain
high effort via explicit contracts \citep{holmstrom1982moral},
relational contracts \citep{levin2003relational} or efficiency wages
\citep{katz1986efficiency}. In each of these theories, the explanation
hinges upon the relationship between the firm and the individual
worker---a worker's co-workers or ``peers,'' if they matter at all,
are relevant only to the extent that they influence the incentives
offered by the firm (e.g., by influencing payoffs in a relative
performance scheme). However, recent empirical research has
highlighted the direct effects that co-workers can have on each others
productivity without intermediation by the firm.\footnote{See, for
  example, \cite{bandiera2010social}, \cite{mas2009peers},
  \cite{falk2006clean}, and \cite{guryan2009peer}.}  There are several
potential channels through which these workplace peer effects could
flow: peers could offer instruction about how to be more productive,
threaten punishment, promise rewards, offer examples of relevant norms
or spur competition. The purpose of this paper is to help clarify
these potential mechanisms through experimentation in a real work
setting.

\subsection{Overview of experiments and findings} 
This paper reports the results of five closely related field
experiments designed to explore the relationships among a worker's
peers, the policies and statements of the firm and a worker's
productivity. Table \ref{overview} provides an overview of each
experimental design and the results. In each experiment, workers from
an online labor market were hired to produce descriptive labels for
photographic images.\footnote{For example, a photograph of a breakfast
  scene might generate the labels ``juice, toast, cereal.''} In each
experiment, all subjects labeled the exact same image which makes
output comparisons across experimental groups meaningful. Before
joining the experiment, would-be workers read a description of the
task, learned the payment and viewed a work sample (i.e., a screen
shot of the image-labeling interface with some number of labels
completed).  If they chose to participate, they labeled one or more
images, depending on the details of the experiment. A worker's output
was simply the number of labels they produced. Because subjects were
not informed they were participating in experiments, the experiments
were ``natural'' field experiments in the \cite{harrison2004field}
taxonomy.

In Experiment A, subjects were randomly assigned to view either a
high-output work sample (with many labels for an image) or a
low-output work sample (with only a few labels for that same image).
All subjects then performed an image-labeling task. All subjects
labeled the same image, making cross-group output comparisons
meaningful. Exposure to the high-output work sample lowered labor
supply on the extensive margin but raised it on the intensive
margin. These two results are important for follow-on experiments,
because they imply that (1) subjects regarded effort as costly and (2)
subjects held the work sample as informative about employer
expectations.

In Experiment B, all subjects viewed the same low-output work sample
from Experiment A and then completed an image-labeling task.  After
completing this task, subjects evaluated the work of another
worker. The evaluated work displayed either high- or low-output, and
subjects were randomly assigned to the work they evaluated. Evaluation
had two parts: each subject was asked to recommend whether or not the
firm should ``approve'' the evaluated work as well as how to split a
bonus with the evaluated worker. The bonus split question created a
contextualized dictator game. The ``approve'' question has a technical
and consequential meaning in the market---when work is not approved,
the worker submitting that work does not get paid and their reputation
suffers.\footnote{A worker's reputation in this market is simply the
  percentage of their submissions that get approved. Buyers can put
  approval percentage screening criteria on their tasks, e.g., only
  allow workers with a 95\% approval rate to complete this task.}
Evaluating low-output work led to greater punishment: subjects viewing
low-output work were far more likely to recommend rejection and
granted smaller bonuses. Regardless of group assignment, highly
productive subjects (as measured by their output on the initial task)
were harsher judges; compared to their low-productivity peers, they
were more likely to recommend rejection and they granted smaller
bonuses.

All subjects in Experiment C were first shown a low productivity work
sample. Subjects then performed an image-labeling task. After
completing the task, subjects next evaluated the work product of
another worker. Unlike in Experiment B, all subjects in C evaluated
the same work. The only experimental manipulation in Experiment C
occurred during the image-labeling phase: subjects were randomly
assigned to either a normal image-labeling interface or to a special
image-labeling interface that generated a pop-up notice after subjects
added their second label. This pop-up notice was designed to modify
subjects' beliefs about employer expectations. The purpose of the
notice was to reduce output without inducing a change in extensive
labor supply.\footnote{Because the pop-up notice appeared after a
  subject had already decided to provide a positive amount of labor,
  it had no effect on the extensive margin.} By changing output
without changing the composition of subjects (i.e., no supply effects
on the extensive margin), it was possible to test the ``innate types''
explanation for the strong relationship between productivity and
punishment found in Experiment B. In this experiment, I found no
evidence that highly productive workers are simply more punishment
prone: workers receiving the pop-up notice reduced their output,
decreased their rejection recommendations and increased their bonuses.

In Experiment D, all subjects were first shown a low productivity work
sample. Next, they performed an initial image labeling task and then
were randomly assigned to evaluate either high- or low-output
work. After this evaluating, subjects performed an additional
image-labeling task. On average, workers that evaluated highly
productive work produced more labels in the follow-on image-labeling
task than workers that evaluated less productive work. These effects
on productivity were strong and easily detectable, but they were not
homogeneous: less productive workers were far less susceptible to the
effects, contra to some findings that peer effects raise the output of
low productivity workers \citep{falk2006clean,mas2009peers}.

In Experiment E all subjects were shown a work sample with exactly $2$
labels and were told that they should produce only $2$ labels. After
performing an initial image-labeling task, subjects evaluated work
that contained either $2$ or $11$ labels. Workers did not treat the
high-effort but non-complying work as worthy of punishment: subjects
were just as likely to recommend approval of the $11$ label work and
granted slightly larger bonus payments. Despite the explicit
statements of employer expectations, exposure to the
high-output/non-complying work had the same effect as exposure to
high-output work in Experiment D: exposed workers raised output, in
many cases beyond the clearly communicated ceiling. However, most
workers complied with the standard initially, and exposure to
complying work seemed to further increase compliance.

\subsection{Implications} 
One explanation for the results across the five experiments is that
workers were uncertain about what constituted an appropriate amount of
output and they use observations from employer-provided work samples
and the output of peers to determine that amount. Because workers find
labeling costly, these beliefs about employer expectations serve as a
constraint in the implicit optimization problem faced by workers.

Learning about employer expectations results in changes not only in a
worker's labor supply, but also in their willingness to punish or
reward their peers. As this learning can come from multiple sources,
perceived employer expectations do not fall wholly under any single
entity's control and can evolve as workers work, observe, and are
observed.

Punishment seems to come easily to many workers. The reasons why
workers punish is unclear, but there are several possible theoretical
explanations. Perhaps the simplest is that workers view themselves as
a monitored agent of the employer, and they make decisions about
acceptance and bonuses according to what they believe will satisfy the
principal. However, workers do not appear to be general-purpose
enforcers of employers' requests---workers punish low
effort. Non-complying but high effort work is treated no differently
vis-\`{a}-vis punishment than complying work. The fact that workers
only appear willing to punish low effort places a constraint on how
firms can make use of worker-driven norm enforcement. For example, it
may be difficult to get workers to substitute easy, correct procedures
for difficult, inefficient procedures. Ironically, the difficulty
itself might make an outdated procedure harder to replace, as workers
who adopt the easier method might be perceived to be shirking.

The finding that exposure to low-output work lowers output, combined
with the finding that low-productivity reduces willingness to punish,
suggests the possibility of an organizational vicious cycle: after
observing idiosyncratically bad work, workers may lower their own
output and punish less in response, in turn reducing other workers'
incentives to be highly productive. This may explain why
organizational leaders often use the language of contagion to describe
morale and so much of management theory focuses on understanding and
influencing organizational culture \citep{schein2004organizational},
rather than, for example, trying to write perfectly complete
employment contracts.

\subsection{Related work} 
Several recent papers examine the effects of peers on workplace
productivity. Perhaps the most illuminating observational evidence
comes from Mas and Moretti (\citeyear{mas2009peers}), who showed that
less productive grocery clerks exhibited greater productivity when
working near highly productive clerks, but only when they were in the
direct view of the highly productive clerks. This finding suggests
that the threat of punishment might partially explain workplace peer
effects. There is much laboratory literature supporting this
punishment-as-peer-effect view, with several studies showing that
workers will readily bear costs and altruistically punish peers that
free-ride in public goods games \citep{fehr2002altruistic,
  fehr2000cooperation}. This ``strong reciprocity''
\citep{carpenter2009strong} is a powerful peer effect, and although
firms are not perfectly analogous to public goods games, the notion of
worker-enforced productivity norms offers a very general potential
solution to the incentive problem of team production.

Guryan, et al. (\citeyear{guryan2009peer}) also use evidence from a
real workplace, albeit an unusual one: they exploited the random
assignment of professional golfers to tournament foursomes to estimate
the effects of each player's peers on the player's own performance. In
contrast to Mas and Moretti, Guryan, et al. found no evidence of peer
effects, providing a useful corrective to hasty or overly broad
generalizations. However, professional golf tournaments are unusual
work environments, in that two common channels for peer effects are
foreclosed: professional golfers know what constitutes good
performance and are unlikely to raise their quality of play in the
``shadow'' of punishment that might be meted out for non-compliance
with productivity norms. In marked contrast with the Mas and Moretti
setting, shirking by a professional golfer imposes a positive
externality on ``co-workers.''

Using a field experiment, Falk and Ichino (\citeyear{falk2006clean})
showed that workers stuffing envelopes in pairs had less variation in
their output levels than synthetically ``paired'' workers constructed
from an experimental group whose members worked alone. While they
cannot estimate the direction of peer effects (i.e., low productivity
affects or is affected by high productivity, or some amalgamation of
effects), their analysis of the output distribution led them to
conclude that it was more likely that less productive workers were
made more productive by working in pairs.

The difference between the Guryan, et al. setting, the Falk and Ichino
setting and that of Mas and Moretti---and the resultant difference in
findings---serves as a justification for the present study, which has
an unusual but highly controllable work context that preserves some of
the common features of work environments, including uncertainty about
norms, costly effort and a task unlikely to inspire much intrinsic
motivation. Unlike the Mas and Moretti setting, however, there are no
overt free-riding externalities (in the check-out line, slacking by
one clerk increases the work load of other clerks). The absence of
direct negative externalities is important, as evidence from such a
setting can provide some sense of how general punishment might operate
in the workplace.  

\subsection{Contribution} 
This paper contributes to the emerging literature on workplace peer
effects. It provides credible evidence of the existence and operation
of peer effects on productivity, which is especially useful given the
lack of concordance between some of the major results in the field and
the inherent difficulty of estimating these kinds of effects
\citep{manski1993}. This evidence is particularly useful because the
scope of possible interpretations is limited, due to the narrow
channel through which peer effects could operate. Observation of work
output was the only ``interaction'' and the payment scheme was not
relative. The punishment component provides additional insight into
the shadow cast by peer-based norm enforcement.

One methodological advance of this paper is that productivity is
measured both before and after exposure to peers. By base-lining prior
output, the conditional nature of peer effects becomes apparent.  For
example, I find that the conditional treatment effects of exposure to
low quality peers differ from the effects found by Falk and Ichino.
They found that ``bad apples far from damaging good apples seem
instead to gain in quality when paired with the latter.''  Mas and
Moretti find a similar result. In the setting examined here, the
traditional bad apples metaphor applied---the bad apples ruined the
good apples, and the good apples did nothing for the bad.
%

\section{Methods and Materials} 
Before describing the experiment results, I first describe the
marketplace where the experiments were conducted, the methodological
issues involved in online experimentation and the actual task
completed by workers and interface used. The experiments were
conducted on Amazon's Mechanical Turk (MTurk), an online labor market
where workers are available to complete small tasks for
payment. Background information on MTurk closely follows
\cite{horton2010labor}. MTurk is one of several online labor markets
that have emerged in recent years \citep{frei2009}. At present, it is
the most amenable to online experimentation.
\begin{table}
\caption{Details of the Experiments \label{overview}}
\begin{center}
\begin{tiny}
    \begin{tabular}{ l | p{3cm} | p{5cm} | p{2cm} | p{2cm} | p{3cm} }
     \small{Exp}. 
     & \small{Question} 
     & \small{Set-up} 
     & \small{Treatment} 
     & \small{Control} 
     & \small{Result} \\ 
    \hline A 
      & Can employers convey productivity expectations? 
      & Subjects viewed an employer-provided work sample, 
        then chose how many labels to produce (if any). Work 
        samples differed by experimental group. 
      & HIGH:Subjects viewed high-output work sample (many labels)
      & LOW: Subjects viewed low-output work sample (few labels)
      & HIGH increased labor supply on intensive margin, 
        but decreased it on extensive margin \\
    \hline  B 
      & Do workers punish workers that exhibit low productivity? 
      & Subjects viewed an employer-provided work sample, 
        then chose how many labels to produce. Subjects then 
        evaluated another worker's work product.  
      & GOOD: Subjects evaluated a high-output work sample
      & BAD:  Subjects evaluated a low-output work sample
      & GOOD increased approval recommendations and bonus amounts. 
        Highly productive workers punished more with their evaluations. \\
     \hline C
      & Is the relationship between own-productivity and punishment causal?
      & Subjects viewed an employer-provided work sample, 
        then chose how many labels to produce. Subjects then 
        evaluated another worker's work product.       
      & CURB: Subjects received a notice after two labels 
        saying that 3 labels was probably enough output
      & NONE: Subjects received no notice. 
      & Greatly reduced output in CURB;
        those in CURB more likely to recommend approval 
        and grant larger bonuses \\
      \hline D
      & Does exposure to low-output work affect a worker's productivity?
      & Subjects viewed an employer-provided work sample, 
        then chose how many labels to produce. Subjects then 
        evaluated another worker's work product. Then they labeled
        a second image. 
      & GOOD: Subjects evaluated a high-output work sample
      & BAD:  Subjects evaluated a low-output work sample
      & GOOD raised output on second task;
        effects were stronger for more productive 
        subjects (measured by first task output). \\
      \hline E
      & Are workers susceptible to peer effects 
        in presence of strongly-stated employer expectations? 
        Do they punish high-effort but non-complying work? 
      & Subjects viewed an employer-provided work sample with
        2 labels, then chose how many labels to produce. Subjects
        were told that 2 and only 2 labels should be produced. 
        Subjects then evaluated another worker's work product, 
        then labeled a second image.  
      & OVER: Subjects evaluated a worker producing too many labels
      & OK: Subjects evaluated a worker producing the required 
        number of labels 
      & OVER increased subsequent output beyond ceiling, but did not cause
        more punishment.  
\end{tabular}
\end{tiny}
\end{center}
\end{table}

\subsection{Online experimentation}
In the past few years, researchers in a number of disciplines---with
computer science leading the way---have begun running experiments
online using online labor markets.\footnote{For an overview of online
  labor markets, see \cite{hortonOLM2010}.} Some examples in economics
include \cite{mason2009fip}, \cite{chandler2010} and
\cite{horton2010labor}. Horton, Rand and Zeckhauser
(\citeyear{hortonZeck2010}) argue that online experiments can offer a
high degree of both internal and external validity. Despite their
advantages online experiments can also be harder to control compared
to conventional laboratory experiments. However, they are generally
easier to control than conventional field experiments. Because
subjects may quit at any time, the biggest threat to valid inference
is non-random attrition. In Experiment A, quitting was actually a
useful outcome to observe, as the experiment focused on labor supply
on both the intensive and extensive margins. In the other four
experiments, by design, essentially all attrition occurred before
subjects experienced any treatment-specific differences. For
Experiments B-E, only subjects that completed the initial
image-labeling task were included in the sample (with others dropped),
but this creates no sampling bias, since all subjects made their
initial output decisions before being exposed to any experimental
group-specific treatments.

\subsection{Amazon's Mechanical Turk} 
Amazon's Mechanical Turk is an online labor market where workers are
available to perform small jobs called ``Human Intelligence Tasks''
(HITs) for buyers, who, in the parlance of MTurk, are called
``requesters.''  HITs vary, but most are small, simple tasks that are
difficult for computers but relatively easy for humans to
perform. Common tasks include transcribing audio clips, classifying
and tagging images, reviewing documents and checking websites for
pornographic content. When posting a HIT, a requester describes the
task, creates a user interface, establishes a piece-rate payment,
specifies worker qualifications, and sets the number of times each HIT
may be performed.

In order to become an MTurk worker, a person must create an MTurk
account and provide a bank account number to Amazon. Workers are only
allowed to have one account, and Amazon uses several technical and
legal means to enforce this restriction. Once they are members,
workers are able to observe the collection of HITs available to them
and, in most cases, view a sample of the required work.  They can work
on any task for which they are qualified and can begin work
immediately after accepting a HIT.

Once a worker completes a HIT, the work product is submitted to the
requester for review. The requester decides whether or not to
``approve'' it. If approved, the worker is paid the piece rate. The
worker is also paid if the requester does not review and approve the
work within a specified amount of time. Solely at their discretion,
requesters may ``reject'' work, in which case the worker is not
paid. The ability of requesters to reject work creates consequences
for providing work that does not meet employer expectations---a
feature critical to the experiments conducted. Requesters may also
elect to pay bonuses, which makes it easy to tailor payments to
individual workers based on their performance within a nominally
piece-rate HIT.

MTurk workers appear to be split approximately evenly between the US
and India. Most report that they participate to earn money and
generally view employers online as having the same level of
trustworthiness as offline, traditional employers
\citep{horton2010condition}. For the demographics of the MTurk
population, see \cite{ipeirotis2010}.

\subsection{Task and interface}
In each of the experiments, subjects were asked to label images. The
images themselves were selected from the photo sharing website
Flickr.\footnote{The images each had a Creative Commons license and
  were chosen because they were conducive to labeling (e.g., photos
  depicting elaborate meals with many easily recognizable food
  types).} Image-labeling is a very common ``human computation'' task
because labels are needed to make images searchable, but computers do
a poor job of identifying objects in images \citep{von2004labeling,
  huang2010toward}. The interface itself was created in Limesurvey, an
open-source survey platform.\footnote{The interface for adding labels
  was written in JavaScript as an add-on module to Limesurvey.} In
order to add labels, subjects had to click a button labeled ``Add a
label.''  A screen shot of the interface can be seen in Figure
\ref{fig:work.samples}.  Clicking the button caused a new blank text
field to be added to the survey. When they finished adding labels
subjects clicked a button labeled ``Submit labels,'' which saved
within the survey all of the labels generated and the time spent
adding labels. No attempt was made to adjudicate the quality of the
labels---if the subject started to add another label, this was
recorded as an additional unit of output.

For the evaluation task, each subject viewed a screen shot of another
worker's work product. As with the initial image-labeling task, the
evaluation task (and the potential bonus) was likely perceived as
unextraordinary. On MTurk, it is very common to have workers evaluate
the work of other MTurk workers. A frequently used solution to the
problem of spam submissions (which occur often when buyers post many
tasks---see \cite{ipeirotis2010quality}) is to have workers vote on
the work product of other workers. It is also very common to use
bonuses to motivate performance.

It is important to note that all the subjects in an experiment labeled
the same image, regardless of their group assignment. For example, in
Experiment A, all subjects labeled an image showing a collection of
mechanics tools. What varied across experimental groups were factors
like the provided work sample, work instructions or the demonstrated
productivity of the worker they were asked to evaluate. Because the
images were the same across groups, output levels are comparable, and
differences in output can be attributed causally to whatever factor
was manipulated in that experiment.

\subsection{Demographic survey}
In each of the five experiments, subjects answered a short demographic
survey before beginning work. The survey was identical in each
experiment. Subjects were asked to report their gender, country
(choices were US, India or some other country) and whether they use
MTurk primarily in order to make money, learn new skills or have
fun. There were small differences in the reported covariates across
experiments, with most of the difference probably driven by
differences in when experiments were launched. 

Although one might think the survey would rise suspicions that the
task was an experiment, I view this as unlikely. Asking workers for basic
demographic information is fairly common in the market, as requesters
frequently use location and formal qualifications to screen workers. In
fact, place- and reputation-based screening is built into the system,
making it even more likely that workers view the demographic questions
as an employer work-around to further expand the ability to screen or
algorithmically adjudicate response quality. With some noted
exceptions, the demographic information had little predictive power
and only marginally improved the precision in the regressions, so they
were not included.

\section{Experiment A: Perceived employer expectations} 
\begin{figure}
  \centering 
   \subfloat[][$HIGH$ ($y=9$)]{  \includegraphics[scale=.25]{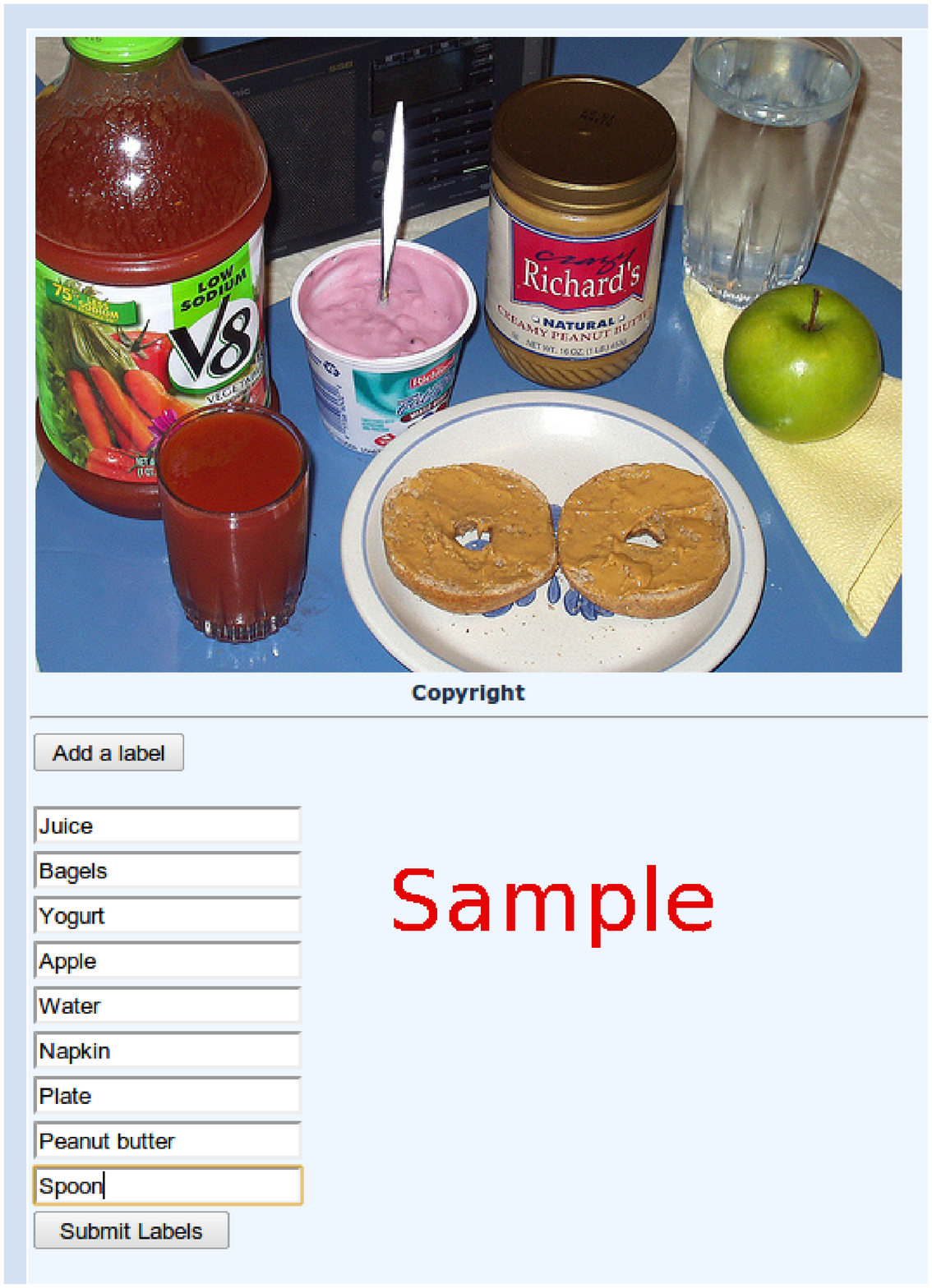}}
   \subfloat[][$LOW$ ($y=2$)]{  \includegraphics[scale=.25]{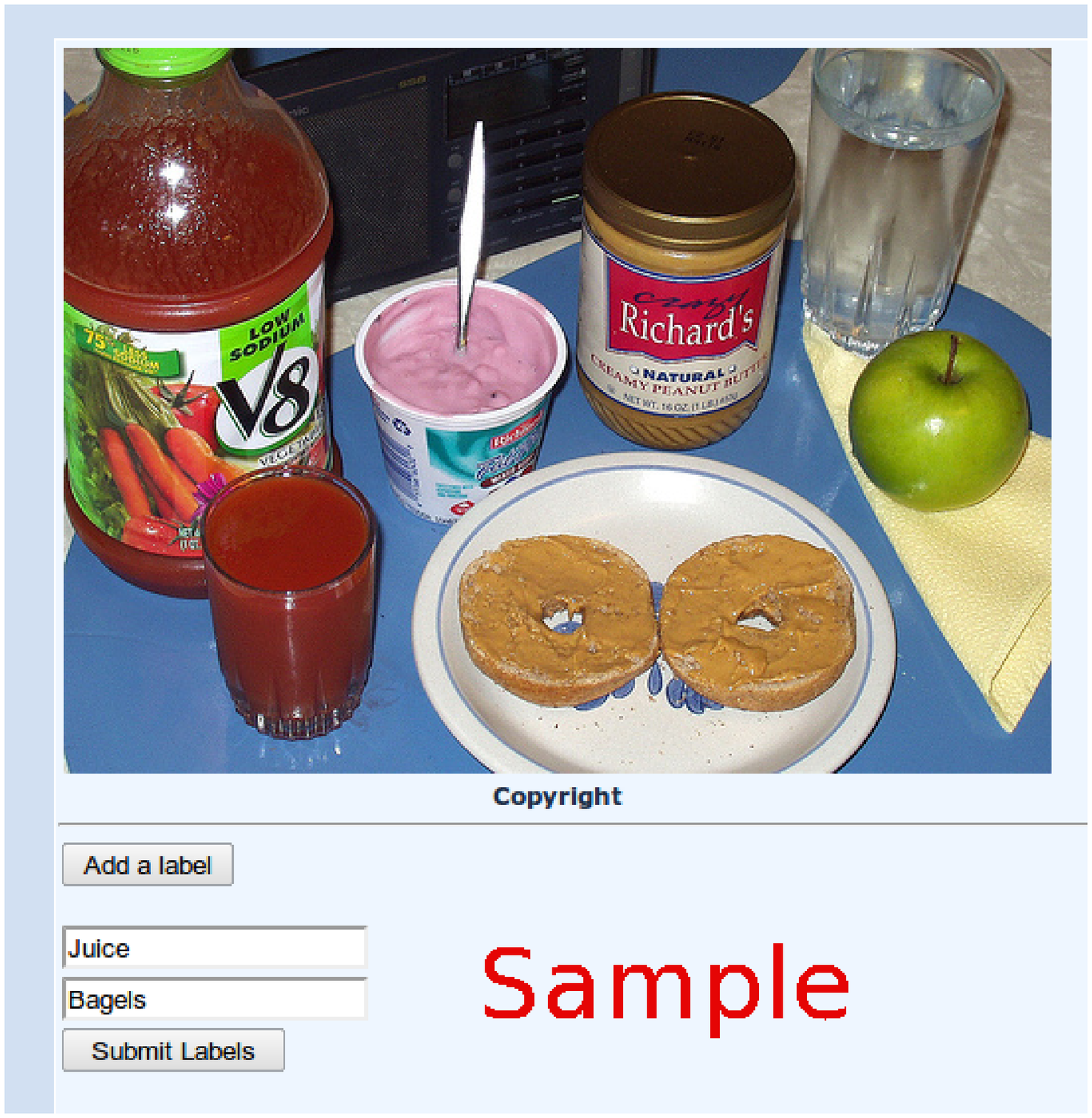}}
  \caption{Work samples shown to workers prior to task acceptance in
    Experiment A.\label{fig:work.samples}}
\end{figure}

Experiment A investigates whether a ``firm'' in this market can
influence employee expectations about productivity. For the image
labeling task, a simple way to convey expectations is to show a work
sample. In this experiment, workers were assigned to one of two
experimental groups: $HIGH$, in which the work sample showed $9$
labels, and $LOW$, in which the work sample showed $2$
labels.\footnote{Throughout the paper, the experimental group names
  will be treated as indicator variables, i.e., $HIGH=1$ is synonymous
  with worker $i$ being assigned to group $HIGH$.}  The work samples
are shown in Figure \ref{fig:work.samples}.  After subjects viewed
their assigned work sample, they chose to either label an image or
exit the experiment, forfeiting payment.


Table \ref{tab:ExpA.ss} shows the summary statistics for the
experiment. The job posting explained that workers would be asked to
do a simple image-labeling task and would be paid 30 cents. The
planned sample size was 100. Subjects not completing the demographic
survey (which occurred prior to group assignment) were dropped from
the sample. In this experiment and all subsequent experiments,
subjects were assigned to groups by stratifying on arrival time (e.g.,
subject 1 was assigned to $HIGH$, subject 2 to $LOW$, subject 3 to
$HIGH$ and so on).

\setkeys{Gin}{width=\imgWs \textwidth}
\setkeys{Gin}{width=\imgW\textwidth}

\begin{table}[h!]
  \begin{center} 
   \caption{Experiment A summary statistics ($n=93$) 
     \label{tab:ExpA.ss}}
\begin{tabular}{lcccccc}
  \toprule 
  \underline{Administrative} \\
    \multicolumn{4}{l}{\hspace{10pt} Launch: Fri Apr 09 21:10:31 GMT 2010} \\
    \multicolumn{4}{l}{\hspace{10pt} Finish:  Sun Apr 11 10:04:40 GMT 2010} \\[5pt] 
  \underline{Survey}    & \underline{FALSE} & \underline{TRUE} & \underline{\% TRUE}   \\
  \hspace{10pt} male                  & 39 & 54 & 58.1  \\
  \hspace{10pt} from India                 & 48 & 45 & 48.4  \\
  \hspace{10pt} from US                    & 58 & 35 & 37.6  \\
  \hspace{10pt} motivated by money                 & 22 & 71 & 76.3  \\[5pt]
  \underline{Treatment Assignment} \\ 
  \hspace{10pt} $HIGH=1$                & 47 & 46 & 49.5  \\[5pt]
 
  \underline{Output} &  \underline{Min} & \underline{.25} & \underline{Med.} & \underline{Mean} & \underline{.75} & \underline{Max}\\ 
 
  Labels produced ($y$) \\
  \hspace{10pt} in HIGH           & 0 & 0 & 1 & 4.609 & 8.5 & 20\\
  \hspace{10pt} in LOW            & 0 & 1 & 2 & 2.638 & 3 & 13\\
  Entry ($y > 0$) \\
  \hspace{10pt} in HIGH           & 0 & 0 & 1 & 0.6957 & 1 & 1\\
  \hspace{10pt} in LOW            & 0 & 1 & 1 & 0.8723 & 1 & 1\\
  Time spent on task (seconds) \\ 
  \hspace{10pt} in HIGH     & 9.157 & 53.27 & 98.13 & 172.9 & 225.4 & 715.5  \\
  \hspace{10pt} in LOW      & 9.563 & 28.61 & 54.9 & 94.85 & 110.4 & 904.3  \\
  \bottomrule 
  \end{tabular}
\end{center} 
The key result from Experiment A can be seen in the differences in
group means in the ``Labels produced'' and the ``Entry'' rows.
\end{table} 

\subsection{Results} 
\begin{figure} 
  \centering
\includegraphics{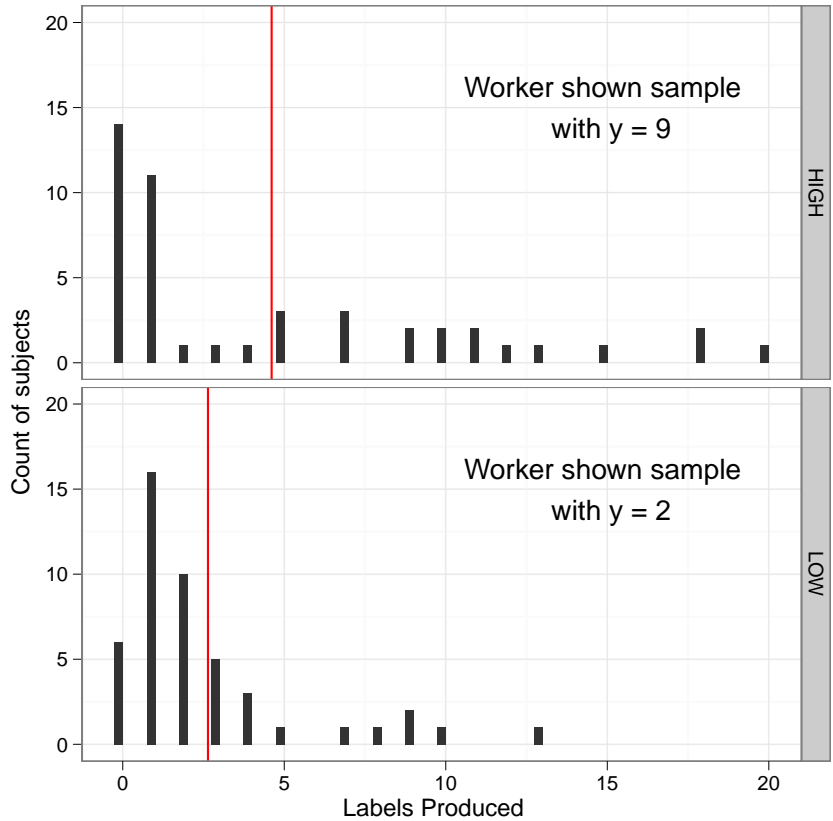}
\caption{Histogram of labels produced by experimental group in
  Experiment A\label{fig:ExpA.output}. The solid vertical line
  indicates the mean, while the shaded band around that line is $2$
  standard errors wide. Subjects in the $HIGH$ group were shown a work
  sample consisting of $9$ labels prior to performing, while subjects
  in $LOW$ were shown a work sample with only $2$ labels. The right
  edge of each bar intersects the x-axis at the corresponding member
  of the support, i.e., the largest output choice in the $HIGH$ panel
  is $y=20$.  This plot and all plots in the document were made using
  the open source R package ggplot2 \citep{wickham2008ggplot2}.}
\end{figure}

The main results from the experiment are displayed in Figure
\ref{fig:ExpA.output}, which shows histograms of output for each
experimental group. Mean output is indicated via a vertical line in
each panel. Subjects in $HIGH$ produced absolutely more output. A
sizable number of subjects in $HIGH$ produced more than $12$ labels,
but only 1 subject in $LOW$ produced more than $12$ labels. However,
the greater productivity in $HIGH$ came at a cost: a larger fraction
of subjects in $HIGH$ elected not to produce any output and quit.

\subsubsection{High employer expectations reduced labor supply on the extensive margin}

When we regress an indicator for any output at all on the treatment
indicator, we have:\footnote{Standard errors are robust and shown under
  the coefficient.}
\begin{align}
  1\{y > 0\}  = \underbrace{-0.177}_{[0.085]}\cdot HIGH +  
\underbrace{0.872}_{[0.050]}
\end{align} 
with $n = 93$ and $R^2 = 0.05$. Subjects assigned to
$HIGH$ were significantly less likely to accept the task compared to
subjects in $LOW$. 

\subsubsection{High employer expectations increased labor supply on the intensive margin}

Despite the much greater number of subjects in $HIGH$ who chose not to
participate (and thus provided $0$ labels), output was unconditionally
higher in $HIGH$ than in $LOW$. Even with the non-participants
included as $y=0$ observations, subjects in $HIGH$ produced, on
average, roughly $2$ more labels per person:
\begin{align}y = \underbrace{1.970}_{[0.956]}\cdot HIGH +  
\underbrace{2.638}_{[0.430]} \end{align}
with $n = 93$ and $R^2 = 0.05$.

\subsection{Discussion} 
Experiment A suggests that workers use the work sample to infer how
much work will be required to meet the employer's expectations and
thus obtain payment. Given that buyers can reject submitted work, the
labor supply results are consistent with workers viewing label
creation as costly. Some of the subjects decided that the costs of
meeting the perceived requirements for the $HIGH$ group were too high
and chose to exit. Those subjects that stayed worked to the higher
perceived standard and completed the task. The increase in output in
$HIGH$ was the result of some combination of selection and greater
effort. Because unconditional output rose significantly in $HIGH$, we
know that selection alone cannot explain the increase in output.

The results highlight the trade-off firms might face when
communicating standards to employees. Claiming to have high standards
may be counterproductive, depending on the nature of the firm's demand
for labor. In our particular image-labeling application, conveying
high standards via the highly productive work sample was efficient
only if the goal was to minimize the per-label price, but it is easy
to imagine scenarios where this is not the objective. For tasks like
image-labeling, obtaining a large and diverse pool of workers---each
contributing a relatively small amount of outut---may be more useful
than obtaining lots of output from small number of workers , in which
case the high standards conveyed to $HIGH$ would have been
undesirable.

\section{Experiment B: Punishment and peer output}

\begin{figure}
  \centering 
   \subfloat[][$GOOD$ ($y=12$)]{  \includegraphics[scale=.25]{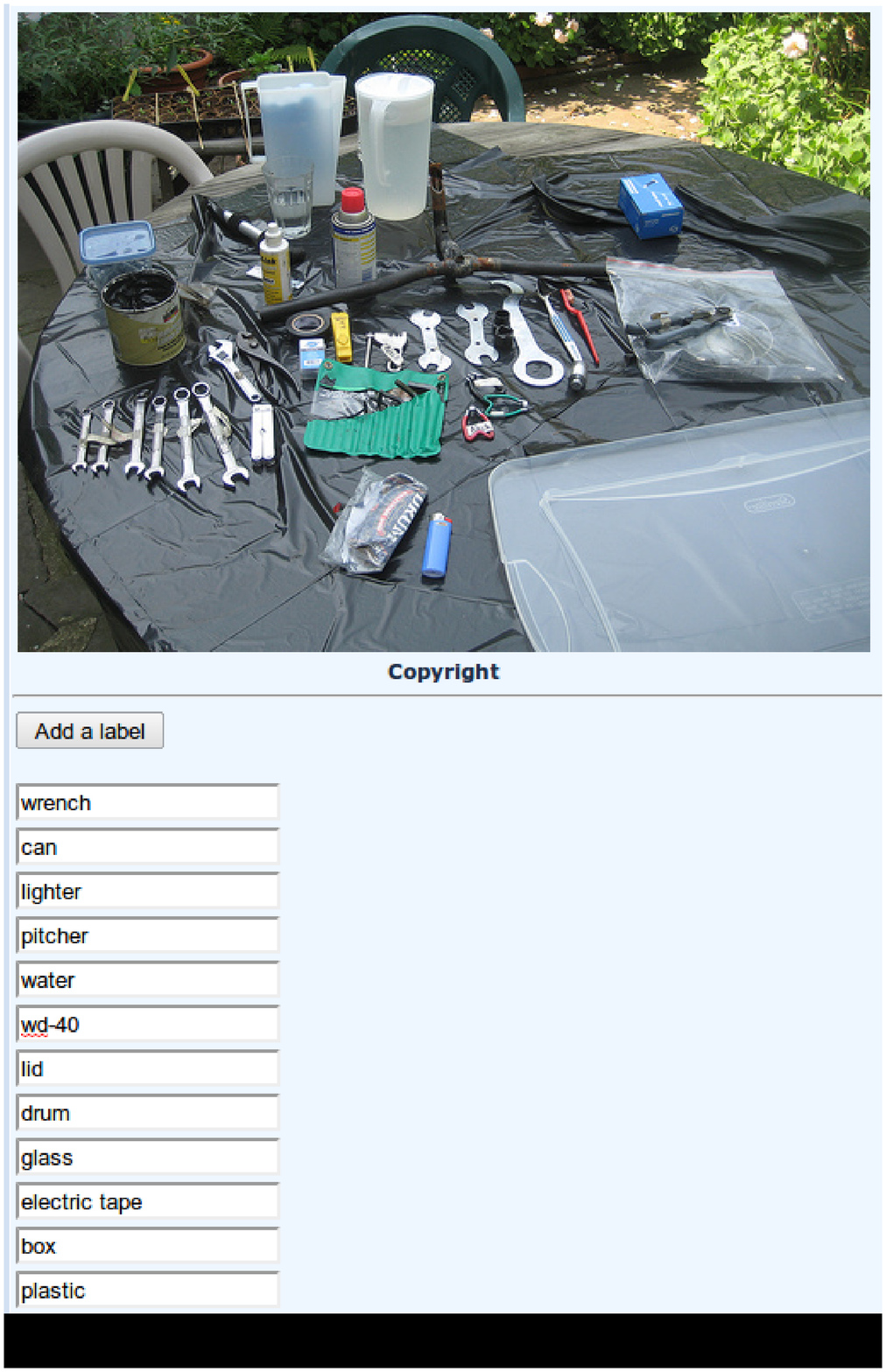}}
    \subfloat[][$BAD$ ($y=1$)]{  \includegraphics[scale=.25]{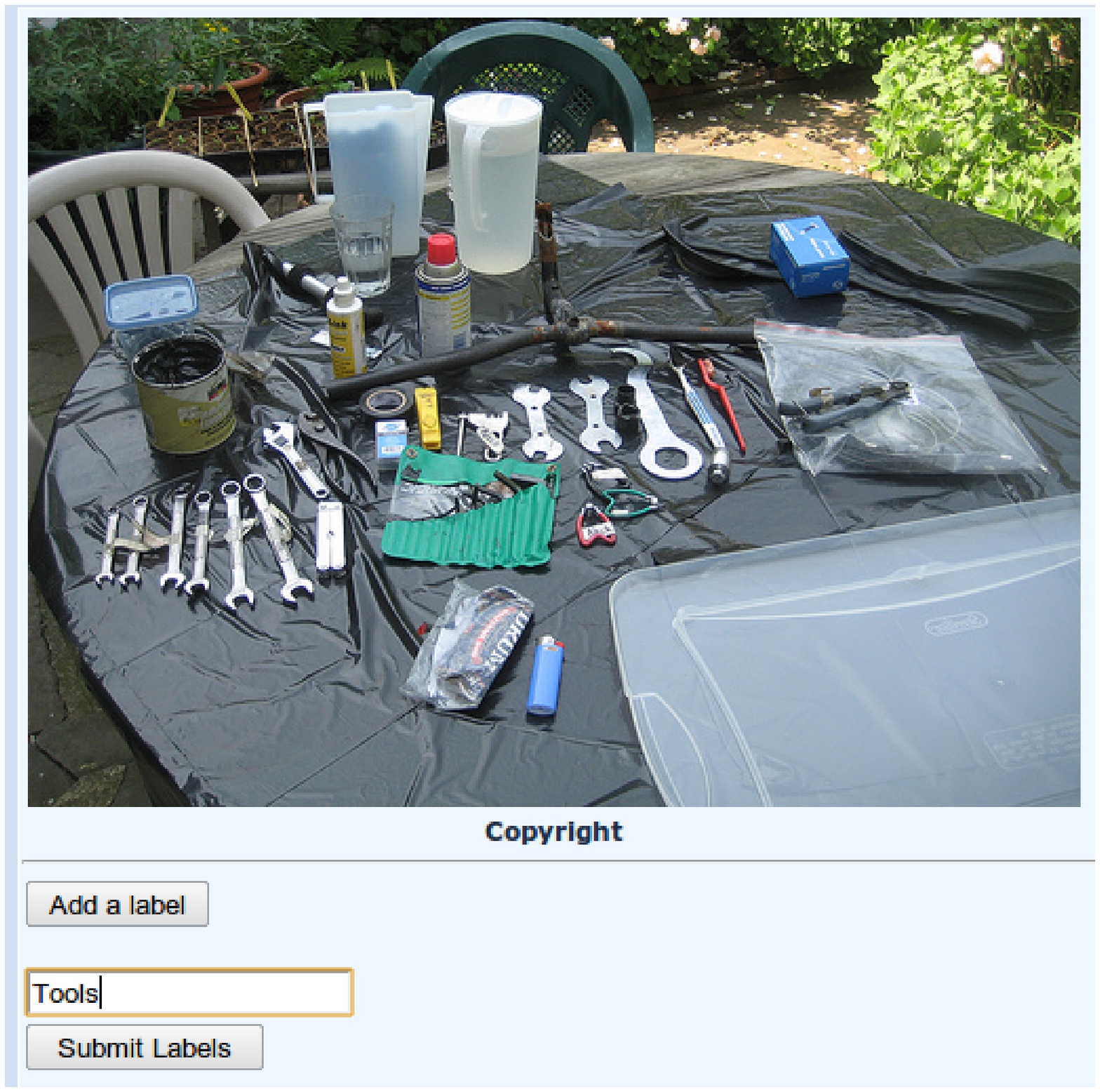}}
  \caption{Work evaluated by subjects in Experiment B \label{fig:work.samples2}}
\end{figure}

Experiment B investigated the conditions under which workers reward or
punish their fellow workers on the basis of their co-workers
output. Subjects in the Experiment first completed an image-labeling
task and then were randomly assigned to either the $GOOD$ or the $BAD$
experimental group. The $GOOD$ subjects inspected the output of a
worker from Experiment A that produced 12 unique labels while $BAD$
subjects inspected the output of a worker that produced only $1$
unique label. The output samples of the evaluated workers are shown in
Figure \ref{fig:work.samples2}. Subjects were then asked to (1) give a
recommendation as to whether or not the work inspected should be
approved and (2) decide how to split a $9$ cent bonus with the
evaluated worker.  Specifically, subjects were asked, ``Should we
approve this work?'' and had to answer ``yes'' or ``no.''  Both
questions were asked on the same survey page, and subjects could
answer them in either order, though the approval question was first on
the page. In the regressions that follow, $approve=1$ indicates that a
subject recommended approval. For the contextualized dictator game,
subjects were told:
\begin{quote}
 ``We want to determine how good this work is. We would like you to
  decide, based on your work and the quality of the other work, how to
  split a 9 cent bonus.''
  \end{quote} 
Subjects selected an answer from a list of $9$ options of the form ``X
cents for them, $9 - X$ cents for me,'' with X ranging from $0$ to $9$
($9$ cents was chosen as the endowment to reduce the salience of the
focal point 50-50 split). At the end of the experiment, we implemented
all choices, with bonuses paid to the evaluating subjects and to the
two lucky subjects responsible for the $GOOD$ and $BAD$ evaluated work
samples. In the regressions that follow, the amount transferred to the
evaluated subject is represented by $bonus$, with $bonus \in
[0,9]$.


The MTurk job posting for Experiment B was nearly identical to the
posting for Experiment A, except that potential subjects were told
that they would be evaluating the work of another worker.  Before
accepting the task, all subjects were shown the $HIGH$ work sample
used in Experiment A. Because of the additional evaluation work, the
participation payment was raised from 30 cents to 40 cents. The
requested sample size was also increased to 200. Table
\ref{tab:ExpB.ss} reports the summary statistics for Experiment B.

\setkeys{Gin}{width=\imgWs \textwidth}

\begin{table}[h!]
 \begin{center} 
   \caption{Experiment B summary statistics ($n=167$) \label{tab:ExpB.ss}}
\begin{tabular}{lcccccc}
  \toprule 
  \underline{Administrative} \\
    \multicolumn{4}{l}{\hspace{10pt} Launch:  Sun Apr 11 04:36:25 GMT 2010} \\
    \multicolumn{4}{l}{\hspace{10pt} Finish:  Mon Apr 12 12:39:20 GMT 2010} \\[5pt]
  \underline{Survey}    & \underline{FALSE} & \underline{TRUE} & \underline{\% TRUE}   \\
  \hspace{10pt} male                  & 70 & 97 & 58.1  \\
  \hspace{10pt} from India                 & 92 & 75 & 44.9  \\
  \hspace{10pt} from US                    & 118 & 49 & 29.3  \\
  \hspace{10pt} motivated by money                 & 41 & 126 & 75.4  \\[5pt]
  \underline{Treatment Assignment} \\ 
  \hspace{10pt} $GOOD=1$                & 82 & 85 & 50.9  \\[5pt]
 
 \underline{Recommended firm approve work}   \\
  \hspace{10pt} in GOOD           & 7 & 78 & 91.8\\
  \hspace{10pt} in BAD           & 41 & 41 & 50\\[5pt]
  
  \underline{Bonus to evaluated worker} &  \underline{Min} & \underline{.25} & \underline{Med.} & \underline{Mean} & \underline{.75} & \underline{Max}\\ 
  \hspace{10pt} in GOOD     & 0 & 4 & 5 & 4.929 & 6 & 9  \\
  \hspace{10pt} in BAD      & 0 & 1 & 3 & 3.488 & 5 & 9  \\
  \bottomrule 
  \end{tabular}
\end{center} 
\emph{Notes:} Overlap in subjects across experiments was $|A \cap
B|=15$.  Subjects in $GOOD$ evaluated work displaying 12 labels, while
subjects in $BAD$ evaluated work displaying only $1$ label. The key
results from the experiments can be seen in the group proportion
differences in the ``Recommended firm approve work'' rows (in the
\%TRUE column) and the group mean differences in the ``Bonus to
evaluated worker'' rows. 
\end{table}

\subsection{Results} 
\begin{figure} 
  \centering
\includegraphics{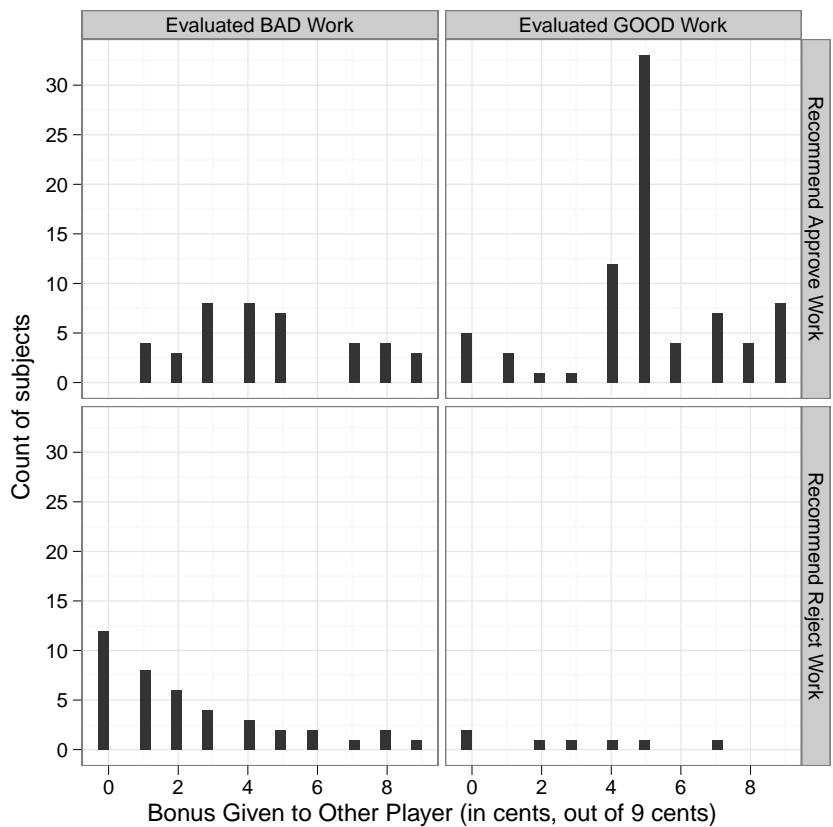}
\caption{Experiment B, bonus allocation by treatment group and
  accept/reject recommendation. Subjects in $BAD$ evaluated work with
  $1$ generic label, while subjects in $GOOD$ evaluated work with $12$
  specific and appropriate labels.\label{fig:Exp.b.a.hist}}
\end{figure}

The results from Experiment B can be seen in Figure
\ref{fig:Exp.b.a.hist}, which contains 4 histograms, each showing the
allocation of the $9$-cent bonus. The plots are faceted by
experimental group (row) and subject recommendation regarding approval
(column). We can see that subjects in $GOOD$ were very unlikely to
recommend rejection, whereas recommendations for rejection were fairly
common among subjects assigned to $BAD$. Few subjects in either group
played the rational strategy of transferring $0$ cents, except among
those subjects in $BAD$ that recommended rejection. For the
$BAD$/reject subjects, the modal transfer was $0$ cents. Most $GOOD$
subjects, as well as a large number of subjects who recommended
approval despite being in $BAD$, chose a more or less equitable split
of $4$ or $5$ cents.

One result to note in Figure \ref{fig:Exp.b.a.hist} in the
$GOOD$/approval quadrant is how generous this distribution is compared
to the usual results of the dictator game. In most laboratory dictator
games, transfers of $0\%$ or $50\%$ of the endowment are common, with
very few subjects transferring more than $50\%$
\citep{engel2010dictator}. Yet a clear majority of subjects in $GOOD$
transferred amounts greater than $50\%$ of the endowment. This
difference is probably due to the very low stakes used in the
experiment.

\setkeys{Gin}{width=\imgW\textwidth}

\subsubsection{Workers more likely to advocate no pay for low productivity work} 

Regressing an indicator for approval on the treatment indicator, we
have:
\begin{align}
  approve = \underbrace{0.418}_{[0.064]}\cdot GOOD + 
\underbrace{0.500}_{[0.056]}
\end{align} with $n =
167$ and $R^2 = 0.21$. Confirming what was
evident graphically, subjects in $GOOD$ were far more likely to
recommend approval.

\subsubsection{Workers rewarded good work with generous bonuses} 

Subjects were more generous to their highly productive peers.
Regressing the amount transferred on the treatment indicator, we have:
\begin{align}
  bonus =  \underbrace{1.442}_{[0.393]}\cdot GOOD + 
  \underbrace{3.488}_{[0.298]} 
\end{align}
with $n = 167$ and $R^2 = 0.08$. 

\subsubsection{Highly productive workers less generous and more likely to advocate rejection}

There is a strong negative correlation between a subject's own
productivity and their generosity in the dictator game:

\begin{align}bonus = \underbrace{-0.259}_{[0.066]} \cdot y + 
  \underbrace{5.376}_{[0.365]} \end{align} 
with $n = 167$ and $R^2 = 0.07$. The
effect also appears strongly in the approval recommendation:
\begin{align}
  approve = \underbrace{-0.055}_{[0.011]} \cdot y + 
  \underbrace{0.960}_{[0.049]} \end{align} 
with $n = 167$ and $R^2 = 0.11$. 

\subsection{Discussion} 
The perceived quality of the evaluated work had a strong causal effect
on measures of both punishment and generosity. Subjects who evaluated
low-output work were far more likely to recommend rejection and
transfer smaller amounts of money in the dictator game.\footnote{Greg
  Little et al. (\citeyear{little2010exploring}) find that when MTurk
  workers evaluate the work product of others, they often use readily
  available metrics such as quantity rather than quality.} The
simplest explanation may be that workers believe that their
evaluations themselves may be spot-checked, and thus it is reasonable
for them to adopt whatever they believe would be the principal's
view. In other words, they reject poor work because they believe that
the employer/requester is likely to be believe it is bad. What is less
clear is why highly productive workers are more likely to reject
work. There are at least three possible explanations:

\begin{itemize} 
  \item There are different ``types'' of workers, and highly
    productive types (measured by producing high output on the initial
    image-labeling task) have a taste for punishment.
  \item Workers idiosyncratically differ in the perception of the
    prevailing productivity norm, and these differences in norm
    perception determine both output choices and norm enforcement.
   \item Workers are inequity-averse \citep{fehr1999theory}, and
     because they regard productivity as costly, they punish low
     effort workers and reward high effort workers as a way to
     equalize outcomes.
\end{itemize} 

\section{Experiment C: Productivity and punishment}
By definition, one cannot experimentally manipulate ``innate types.''
However, if manipulations of productivity cause a change in
willingness to punish, then the innate types hypothesis is
untenable. To distinguish inequity aversion from norm enforcement, one
would need a way of manipulating a worker's experienced productivity
without changing either their perceptions of the prevailing norm or
their perceptions of each party's payoffs. In the parlance of the
treatment effects literature, the exclusion restriction would need to
be satisfied while not creating any non-random attrition. Given our
imperfect understanding of how workers make decisions about both labor
supply and generosity, it is unlikely that the exclusion restriction
can be credibly satisfied. Nevertheless, ruling out the innate types
hypothesis is still worthwhile, which is the goal in Experiment C.

To illustrate the challenge of distinguishing among the hypotheses,
consider the implications of using the setup of Experiment A to induce
changes in labor supply. In Experiment A we were able to change output
by altering the work sample shown to workers before they accepted the
task. We did not measure follow-on output in a second image-labeling
task in Experiment A nor did we have them evaluate others work, but if
we had, at least two problems would arise. First, the manipulation
would have an effect on both the intensive margin and the extensive
margin. Second, subjects who quit in the first stage would not record
their choices in the dictator game or their answer to the
accept/reject question.

The fundamental problem is that any intervention that changes worker
pre-uptake beliefs about the work required---and hence the labor
supply on the extensive margin---is likely to create a missing data
problem. For this reason, the intervention used in Experiment C
changes worker productivity \emph{after} subjects have already begun
working. The design, as well as the failed pilots, will be discussed
in detail, but the key point is that the intervention successfully
manipulated output on the intensive margin and yet did not cause any
across-group differences on the extensive margin (i.e., lead to
differences in group composition). However, it likely did affect
perceived deservingness or inequity, making it impossible to decide
between the two hypotheses related to these concepts.

\subsection{Pilot experiments} 
Prior to running Experiment C, two failed pilot experiments were
conducted. In the first pilot, half of the subjects were assigned to
work with an interface containing a hidden ``bug'' that introduced a
$1.2$-second delay in the software execution after each label was
added; those in the control faced no delay. This pilot failed because
it did not generate a ``first stage'' of reduced output---although
workers in the ``slow'' treatment took longer, they did not produce
any less output. This outcome is consistent with other findings that
workers on MTurk appear insensitive to small time differences
\citep{horton2010labor}.

In the second pilot, subjects in the ``slow'' treatment received a
pop-up box containing the text ``Thank you! Three is probably
enough.''  after they had added a second label but before they added a
third label. Although this treatment did have a large effect on worker
productivity, other aspects of the design of the experiment generated
little variation in generosity.  The first problem was that in order
to obtain a larger sample, the $LOW$ work sample from Experiment A was
used, thereby compressing the productivity distribution (as in
Experiment A). The second problem was that workers evaluated work with
$3$ good labels. As a result, regardless of their treatment
assignment, many subjects chose the pseudo-50-50 (i.e., chose either
the 4 or 5 cent transfer) split and recommended approval, providing
little useful variation in measurements of punishment and generosity.

\subsection{Actual experiment}
After the two failed pilots, the second pop-up notice experiment was
relaunched, albeit with two modifications. The $HIGH$ $9$-label work
sample from Experiment A served as the sample, and the evaluated work
was particularly bad---the evaluated worker provided only $1$ generic
label for an item-rich photograph. Subjects were assigned to either
$NONE$, in which they were given no notice, or $CURB$, in which they
received the pop-up notice after completing a second label. Because
two-stage least squares would be used in the data analysis, the total
sample size was increased to $300$. Payment was 30 cents. Table
\ref{tab:ExpC.ss} reports the summary statistics for experiment.



\begin{table}[h!]
  \begin{center}
   \caption{Experiment C summary statistics ($n=273$)\label{tab:ExpC.ss}}
\begin{tabular}{lcccccc}
  \toprule 
 \underline{Administrative} \\
  \multicolumn{4}{l}{\hspace{10pt} Launch:   Sun Apr 18 19:36:44 GMT 2010} \\
  \multicolumn{4}{l}{\hspace{10pt} Finish:   Wed Apr 21 05:18:02 GMT 2010} \\[5pt]
  \underline{Survey}    & \underline{FALSE} & \underline{TRUE} & \underline{\% TRUE}   \\
  \hspace{10pt} male                  & 117 & 156 & 57.1  \\
  \hspace{10pt} from India                 & 175 & 98 & 35.9  \\
  \hspace{10pt} from US                    & 162 & 111 & 40.7  \\
  \hspace{10pt} motivated by money                 & 80 & 193 & 70.7  \\[5pt]
  \underline{Treatment Assignment} \\ 
  \hspace{10pt} $CURB=1$                & 133 & 140 & 51.3  \\[5pt]
 
 \underline{Recommended we approve work?}   \\
  \hspace{10pt} in CURB           & 46 & 94 & 67.1\\
  \hspace{10pt} in NONE          & 57 & 76 & 57.1\\[5pt]
  
  \underline{Labels produced ($y$)} &  \underline{Min} & \underline{.25} & \underline{Med.} & \underline{Mean} & \underline{.75} & \underline{Max}\\ 
 
   \hspace{10pt} in CURB           & 1 & 2 & 3 & 2.979 & 4 & 10\\
  \hspace{10pt} in NONE            & 1 & 1 & 5 & 6.15 & 9 & 26\\
  \underline{Bonus to evaluated worker} \\
  \hspace{10pt} in CURB     & 0 & 2 & 4 & 3.871 & 5 & 9  \\
  \hspace{10pt} in NONE      & 0 & 1 & 3 & 3.075 & 5 & 9  \\  
  \bottomrule 
  \end{tabular}
\end{center} 
\emph{Notes:} Overlap in subjects across experiments was $|C \cap B| =
20$, $|C \cap A \cap \overline{B}|=20$. Subjects in $CURB$ reduced an
output-reducing notification after adding a second label. The key
results from the experiment can be seen in the mean differences across
groups (in the ``Bonus to evaluated worker'' rows and the ``Labels
produced'' rows).
\end{table}

\subsection{Results} 
\begin{figure} 
  \centering
\includegraphics{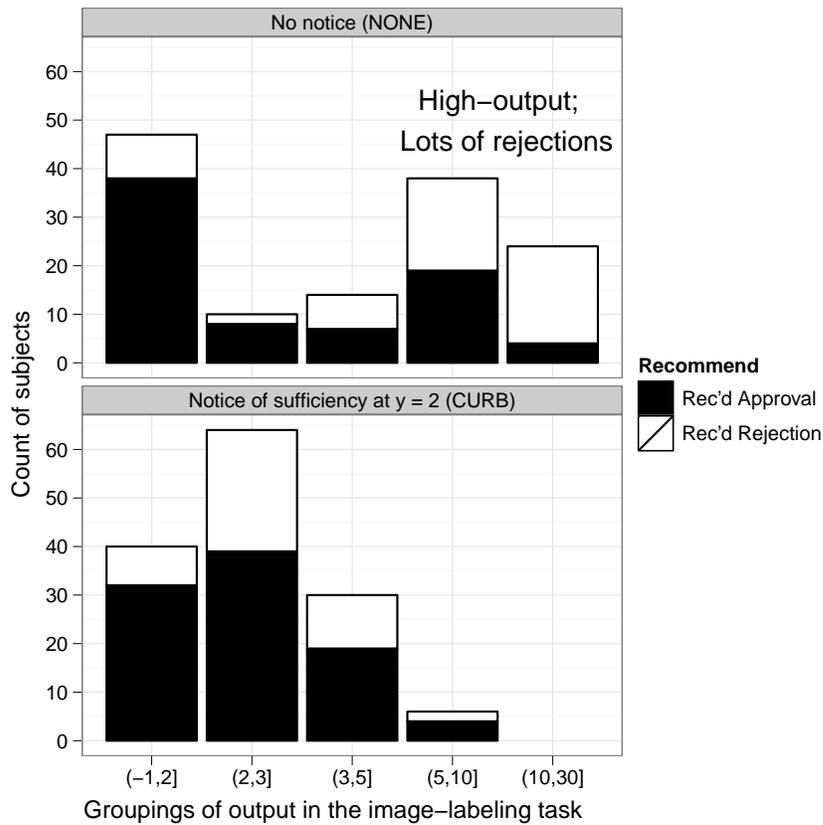}
\caption{Output and punishment in Experiment
  C\label{fig:ExpC.y.a}. This bar chart shows the amount of output in
  different output bands and the percentage of subjects in that band
  recommending approval or rejection. Note that a disproportionate
  number of subjects not receiving the output-curtailing message
  (subjects in $NONE$) produced relatively high levels of output and
  that subjects in that high-output band were much more likely to
  recommend rejection than low-output subjects.}
\end{figure}

Worker output at the image-labeling task is binned and then plotted as
two bar charts in Figure \ref{fig:ExpC.y.a}. The top panel shows the
bar chart for subjects in $NONE$, while the bottom shows $CURB$.  The
bars themselves are filled in with the proportion of subjects in that
band/group recommending rejection or acceptance of the evaluated
work. Several features of the data are readily apparent. First,
assignment to $CURB$ dramatically reduced output: a large numbers of
subjects in $NONE$ produced $y \in (5,10]$ or $y \in (10,30]$. By
    comparison, less than $10$ subjects total in $CURB$ produced this
    much.  Second, subjects choosing high levels of productivity in
    $NONE$, they were far more likely to recommend rejection. Although
    bonus allocation is not shown in the figure, this same pattern
    appears---subjects with reduced productivity (those in $CURB$) were
    more generous in their allocation of the $9$ cents.

\subsubsection{Workers with reduced output more generous}
There is a strong negative correlation between the amount transferred
to the other player in the dictator game and a subject's own output in the image-labeling task, as shown by:
\begin{align}
  bonus = \underbrace{-0.207}_{[0.039]} \cdot y +
  \underbrace{4.418}_{[0.223]} \end{align} with
$n = 273$ and $R^2 = 0.12$. Assignment to
$CURB$ had a strong, negative effect on worker output, as shown by:
\begin{align}
  y = \underbrace{-3.172}_{[0.490]}\cdot CURB 
  +  \underbrace{6.150}_{[0.470]} 
\end{align}
with $n = 273$ and $R^2 = 0.14$. The
F-statistic for the model is 43.982. The two-stage least
squares estimate of the effect of output on transfers in the dictator
game is:
\begin{align}
  bonus = \underbrace{-0.251}_{[0.090]} \cdot y +
  \underbrace{4.619}_{[0.433]} \end{align}
There is no detectable difference between the OLS estimate of the
effect of productivity on bonuses and the two stage least-squares
estimate.

\subsubsection{Workers with reduced output less likely to punish}

As in Experiment B, highly productive subjects were more likely to
recommend that we not approve the evaluated workers' work:
\begin{align}
  approve = \underbrace{-0.042}_{[0.005]}\cdot y + 
\underbrace{0.811}_{[0.037]}
\end{align}
with $n = 273$ and $R^2 = 0.13$. The two-stage least squares estimate is:
\begin{align}
  approve = \underbrace{-0.032}_{[0.017]}\cdot y + 
\underbrace{0.765}_{[0.083]}
\end{align}
which is not significantly different from the least-squares estimate.

\subsection{Discussion}
Experiment C rules out the possibility that productivity and
generosity are jointly determined by some innate worker ``type'':
reducing productivity reduced willingness to punish. However, the
experiment does not distinguish between the ``enforced norms''
hypothesis and the ``inequity aversion'' hypothesis. The problem stems
in part from the difficulty in manipulating worker productivity
without also altering workers' perception of the relevant norm. It
seems possible that future research could disentangle these hypothesis
with a suitable experimental design.

However, other experimental results probably push the balance of
evidence towards the enforced norms hypothesis. First, more complex
laboratory games such as those conducted by
\cite{charness2002understanding} show that workers trade off a number
of competing interests when making dictator game allocations and that
preferences are more nuanced than simple inequity
aversion. \cite{list2008examining} make a compelling argument that
what we often interpret as ``social preferences'' in the dictator game
is in fact a desire to be seen as complying with some context-specific
norm.

\section{Experiment D: Peer effects from evaluation}

Experiment A showed that exposure to employer-provided work samples
affected labor supply, presumably by changing worker beliefs about the
employer output expectations. Experiment D tested whether work samples
from peers that are not held up as examples can still influence
productivity. In set-up, Experiment D was similar to Experiment B in
that after an initial task, subjects were assigned to one of two
groups, $GOOD$ and $BAD$. In $GOOD$, subjects evaluated a worker that
produced 11 labels; in $BAD$, subjects evaluated a worker that
produced only 2 labels. Unlike Experiment B, however, subjects
performed an additional image-labeling task after the
evaluation. Table \ref{tab:ExpD.ss} reports the summary statistics for
the experiment. The requested sample size was $300$ and payment was 40
cents.

%

\begin{table}[h!]
  \begin{center}  
   \caption{Experiment D summary statistics ($n=275$)\label{tab:ExpD.ss}}
\begin{tabular}{lcccccc}
 \toprule 
 \underline{Administrative} \\
  \multicolumn{4}{l}{\hspace{10pt} Launch:  Fri Apr 23 18:37:56 GMT 2010} \\
  \multicolumn{4}{l}{\hspace{10pt} Finish:  Wed Apr 28 12:30:17 GMT 2010} \\[5pt]   
  \underline{Survey}    & \underline{FALSE} & \underline{TRUE} & \underline{\% TRUE}   \\
  \hspace{10pt} male                  & 131 & 144 & 52.4  \\
  \hspace{10pt} from India                 & 177 & 98 & 35.6  \\
  \hspace{10pt} from US                    & 154 & 121 & 44  \\
  \hspace{10pt} motivated by money                 & 76 & 199 & 72.4  \\[5pt]
  \underline{Treatment Assignment} \\ 
  \hspace{10pt} $GOOD=1$                & 142 & 133 & 48.4  \\[5pt]
  
  \underline{Labels produced} &  \underline{Min} & \underline{.25} & \underline{Med.} & \underline{Mean} & \underline{.75} & \underline{Max}\\ 
  Initial output, before evaluation ($y_1$) \\
  \hspace{10pt} in GOOD     & 1 & 2 & 5 & 4.759 & 7 & 14  \\
  \hspace{10pt} in BAD      & 1 & 2 & 5 & 4.768 & 7 & 15  \\
  Follow-on output, after evaluation ($y_2$) \\
  \hspace{10pt} in GOOD           & 0 & 4 & 7 & 7.368 & 10 & 23\\
  \hspace{10pt} in BAD            & 0 & 1.25 & 5 & 5.014 & 7 & 16\\
  
  \bottomrule 
  \end{tabular}
\end{center} 
\emph{Notes:} Overlap in subjects across experiments was $|D \cap
C|=26$, $|D \cap (A \cup B) \cap \overline{C}|=37$. Subjects in $GOOD$
evaluated high-output work, while subjects in $BAD$ evaluated
low-output work. The key finding from this experiment was the effect
exposure had on subsequent output. We can see that there were no
differences in output means pre-exposure (``Initial output, before
evaluation'' rows) but a large difference after evaluation
(``Follow-on output, after evaluation'' rows).
\end{table}

\subsection{Results} 
Exposure to the work of a peer strongly affected a subject's
subsequent output. Output following exposure to highly productive
peers was higher than output following exposure to less productive
peers.  The treatment effect is heterogeneous across productivity
distribution for the first task: more productive workers are more
strongly affected by the exposure to peers.

\begin{figure} 
\centering 
\includegraphics{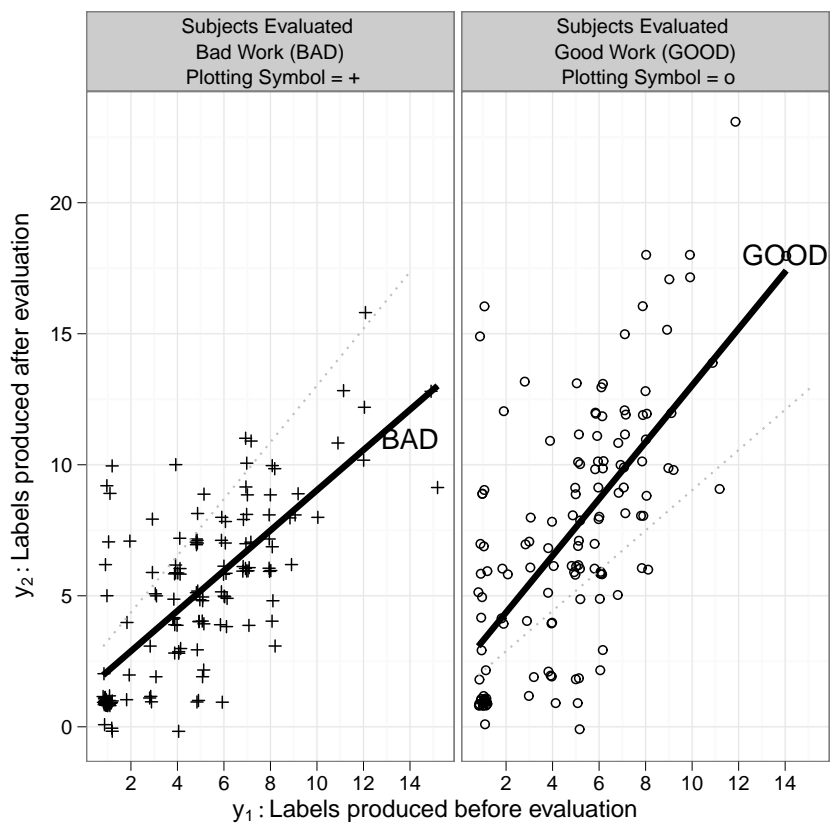}
\caption{Subsequent output ($y_2$) versus initial output ($y_1$), by
  treatment group. All subjects did an identical initial task and
  chose some number of labels to provide (shown on the
  x-axis). Subjects then evaluated another subject's work that
  demonstrated either low productivity ($BAD$, left panel) or high
  productivity ($GOOD$, right panel). All points from each group are
  shown in each panel (as well as the regression line), but the points
  and lines are either black or gray depending on whether they came
  from the experiment group shown in the panel. All output levels are
  randomly perturbed by $\epsilon \sim U[-.2,.2]$ to prevent
  over-plotting (which would occur because only integer output levels
  were possible).
   \label{fig:ExpD.output}}
\end{figure}

Most of the results can be readily seen in Figure
\ref{fig:ExpD.output}, which shows scatter plots of final output,
$y_2$, against initial output, $y_1$. Observations from $BAD$ are
represented in the plot by a ``+'' symbol and those in $GOOD$ by a
``o'' symbol. The two panels contain the scatter plot and regression
line for the respective experimental group, as well as the points from
the other group, lightly plotted.  Because only integer-level outputs
were possible, all points are randomly perturbed to prevent
over-plotting. In the figure, the line for $GOOD$ is both above and
steeper than the regression line for $BAD$, indicating non-constant
effects.

\subsubsection{Exposure to highly productive peers increased productivity}
Subjects that evaluated highly productive work produced considerably more
labels in the follow-on task: 
\begin{align}
  y_2 = \underbrace{0.914}_{[0.071]} \cdot y_1 +  
 \underbrace{2.362}_{[0.373]} \cdot GOOD + 
  \underbrace{0.654}_{[0.398]}  
\end{align}
with $n = 275$ and $R^2 = 0.49$. In
addition to the treatment effect, we can see that initial output was
highly correlated with subsequent output. The effect of $GOOD$ was
not, however, constant across the initial output distribution:
\begin{align}y_2 = \underbrace{0.771}_{[0.073]} \cdot y_1 +  
\underbrace{0.317}_{[0.142]} \cdot y_1GOOD +
\underbrace{0.850}_{[0.799]} \cdot GOOD +
\underbrace{1.337}_{[0.411]}
\end{align} 
with $n =275$ and $R^2 = 0.5$; as $y_1$ increases, the
positive effect of assignment to $GOOD$ on output grows larger.

\subsection{Discussion}
The peer effects detected in Experiment D strongly depend upon a
worker's initial output. There is no immediately obvious reason why
this should be the case, however, studies from other domains have also
found that different types of workers respond differently to
peers. One possible explanation for the pattern is that initially less
productive workers might already have strong beliefs that low
productivity is acceptable. Recall that subjects were exposed to the
$HIGH$ work sample from Experiment A prior to accepting the task, and
yet still chose to produce only $1$ or $2$ labels. Although being in
$GOOD$ and observing highly productive work might still have some
effect on their beliefs, these less productive workers might have
fairly stiff priors regarding what constitutes acceptable work.

Although Experiment D demonstrated that exposure to peer output
affects a worker's own output, it did not explain \emph{why} workers
are influenced by peers.  There are several possible explanations for
why peer effects exist in this setting including fear of punishment,
learning about relevant employer standards and perhaps even an innate
desire to match the performance of peers, regardless of the direct
material payoff.

\section{Experiment E: Peer effects after explicit employer instructions} 
Experiment D demonstrated the existence of productivity peer
effects---even with the minimal ``interaction'' created by evaluation,
yet it did not explain why workers are affected by peers. If peer
effects reflect learning about employer standards, then clear,
strongly stated production standards should ``inoculate'' workers
from learning-driven peer effects. However, if workers fear punishment
by fellow workers or if they have some innate desire to produce as
much as peers, then we should detect peer effects even when standards
are clearly communicated.

In Experiment E, these ideas were tested by providing subjects with
very explicit instructions about productivity expectations and then
exposing subjects to peers. The set-up was almost identical to that of
Experiment D, except that workers were told that they should produce 2
and only 2 labels per image. The requirement of 2 labels was stated
before workers began the task, and was repeated again with each of the
two image-labeling tasks, directly above the image. After performing
the initial task, workers were assigned to one of two groups: $OK$, in
which subjects evaluated a work sample showing $y=2$, and $OVER$, in
which subjects evaluated a work sample showing $y=11$. After
evaluating the work, subjects performed an additional image-labeling
task. Table \ref{tab:ExpE.ss} shows the summary statistics for
the experiment. The requested sample size was $300$ and the payment
was 40 cents.


\begin{figure} 
\centering 
\includegraphics{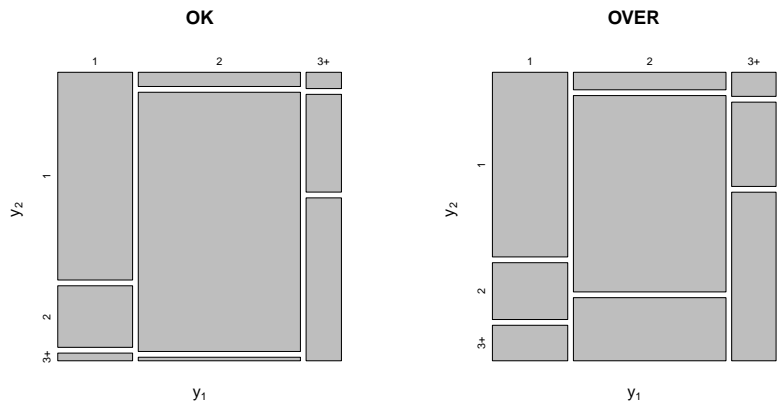}
\caption{Mosaic plots showing the relationship between output in the
  initial task ($y_1$) to output in the follow-on task ($y_2$). In
  Experiment E, all subjects were told to produce $2$ and only $2$
  labels. Subjects in $OK$ then evaluated work showing $2$ labels
  (hence complying with employer expectations), while subjects in
  $OVER$ evaluated work with $11$ labels. In both plots, the unit
  square is split vertically in thirds, in proportion to the number of
  subjects that selected $y_1=1$, $y_1=2$ and $y_1 \ge 3$,
  respectively. Both plots are also split horizontally, in proportion
  to the number of subjects choosing $y_2=1$, $y_2=2$ and $y_2 \ge 3$,
  respectively.  The $(y_1 = 2,y_2 = 2)$ block in $OK$ is taller than
  the corresponding block in $OVER$, indicating that exposure to the
  complying work sample (in $OK$) made initially complying workers
  (i.e., those choosing $y_1=2$) more likely to comply in the follow
  on task (i.e., choosing $y_2=2$).
  \label{fig:ExpE.trans}}
\end{figure}

 \begin{table}[h!]
  \begin{center}
   \caption{Experiment E summary statistics ($n=272$)\label{tab:ExpE.ss}}
\begin{tabular}{lcccccc}
  \toprule 
 \underline{Administrative} \\
  \multicolumn{4}{l}{\hspace{10pt} Launch:  Wed May 19 21:19:42 GMT 2010 } \\
  \multicolumn{4}{l}{\hspace{10pt} Finish:   Sun May 23 15:26:30 GMT 2010} \\[5pt]   
  \underline{Survey}    & \underline{FALSE} & \underline{TRUE} & \underline{\% TRUE}   \\
  \hspace{10pt} male                  & 121 & 151 & 55.5  \\
  \hspace{10pt} from India                 & 136 & 136 & 50  \\
  \hspace{10pt} from US                    & 172 & 100 & 36.8  \\
  \hspace{10pt} motivated by money                 & 62 & 210 & 77.2  \\[5pt]
  \underline{Treatment Assignment} \\ 
  \hspace{10pt} $OK=1$ & 141 & 131 & 48.2 \\[5pt]
  
  \underline{Labels produced} &  \underline{Min} & \underline{.25} & \underline{Med.} & \underline{Mean} & \underline{.75} & \underline{Max}\\ 
  Initial output, before evaluation ($y_1$) \\
  \hspace{10pt} in OK     & 1 & 1 & 2 & 2.038 & 2 & 8  \\
  \hspace{10pt} in OVER      & 1 & 1 & 2 & 2.085 & 2 & 8  \\
  Follow-on output, after evaluation ($y_2$) \\
  \hspace{10pt} in OK           & 1 & 2 & 2 & 1.977 & 2 & 8\\
  \hspace{10pt} in OVER            & 1 & 2 & 2 & 2.667 & 3 & 14\\
  
  \bottomrule 
  \end{tabular}
\end{center} 
\emph{Notes:} Overlap in subjects across experiments was $|E \cap
D|=29$, $|E \cap (A \cup B \cup C) \cap \overline{D}|=42$. All
subjects received explicit instructions to produce 2 labels. After
performing one task, subjects were assigned to $OK$, where they viewed
complying work, or $OVER$, where they viewed high-output/non-complying
work. The key finding from the experiment can be seen in difference in
group means in the ``Follow on output, after evaluation'' rows. 
\end{table} 

\subsection{Results}
The main results of Experiment E are readily apparent graphically, but
the appropriate visualization is somewhat more complex. We would like
to see how the choice of $y_2$ depended on $y_1$ and the assignment to
$OK$ or $OVER$. Unfortunately, a scatter plot would be hard to
interpret, as we expect many subjects to choose $y_1=2$ and $y_2=2$,
per the instructions.  Much information would be lost to
over-plotting. A solution is to use a mosaic plot, which is useful for
displaying relationships between categorical data.

In Figure \ref{fig:ExpE.trans}, the left panel shows a mosaic plot for
$OVER$, while the right panel shows same plot but for $OK$. Each
pair-wise combination of output levels is represented by a rectangle.
Output levels are top-censored, creating three output groups: $y=1$,
$y=2$ and $y \ge 3$ (with $3+$ as a label for the final group).  The
width of each rectangle is proportional to the share of subjects that
chose that respective level of output for $y_1$; the height on each
rectangle is proportional to the number of subjects that chose that
level of output for $y_2$.

Across both groups, most subjects chose $y_1=2$, suggesting that employer
instructions to produce exactly 2 labels were salient. For subjects that
complied initially, exposure to $OK$ was associated with a high level of
compliance on the second task: only a tiny number of subjects
increased or decreased output, as indicated by the very short
$(y_1=2,y_2=1)$ and $(y_1=2,y_2=3+)$ rectangles. However, in $OVER$,
many subjects that chose $y_1=2$ subsequently increased their output
level after evaluating the $11$ label image, as seen by the tall
rectangles associated with $y_2 = 3+$ in $OVER$.

\subsubsection{Most workers compliant with employer output requests} 
In both $OVER$ and $OK$, we can see that $y_1=2$ was by far the most
common output choice. Reassuringly, Figure \ref{fig:ExpE.trans} shows
that the breakdown of $y_1$ appears almost identical across the two
treatments. This is expected since subjects were randomized and the
experience of the groups did not differ until after the first task was
completed. The instruction to produce only two labels appears salient,
especially considering the first stage was identical to that of this
experiment that in Experiment A, $LOW$ (Figure \ref{fig:ExpA.output},
bottom panel), except that no explicit instructions were given, and in
the Experiment A case, $y=2$ was not the modal choice, as it was in
Experiment E.

\subsubsection{Language barriers likely prevented full initial compliance} 

Although not causal, a regression of the compliance indicator on
self-reported country suggests that language barriers might have
limited compliance, with subjects from India significantly less likely
to comply:
\begin{align}
  1\{y_2=2\} = \underbrace{-0.228}_{[0.059]}\cdot INDIA + 
\underbrace{0.691}_{[0.040]}
\end{align} 
with $n = 272$ and $R^2 = 0.05$.

\subsubsection{Evaluating compliant work increased compliance for initially complying workers}
Being assigned to $OK$ strongly increased $y_2$-compliance:
\begin{align} 
  1\{y_2=2\} = \underbrace{0.161}_{[0.059]}\cdot OK + 
\underbrace{0.504}_{[0.042]}\end{align}
with $n =272$ and $R^2 = 0.03$.  This effect is
driven by subjects in $OK$ who initially complied and then continued
to comply on the second image.  This is evidenced by the large and
significant coefficient on the compliance $\times$ assignment
interaction:
\begin{align}1\{y_2=2\} = \underbrace{0.022}_{[0.083]}\cdot OK + 
\underbrace{0.205}_{[0.102]}\cdot \left[OK\times 
1\{y_1=2\} \right] +
\underbrace{0.467}_{[0.076]}\cdot
1\{y_1=2\} +
\underbrace{0.242}_{[0.055]}\end{align} with $n
=272$ and $R^2 = 0.36$. Note that exposure to $OK$
has no effect for originally non-complying workers who chose $y_1 \ne
2$. Also note that initial compliance was strongly predictive
of subsequent compliance.

\subsubsection{Workers did not punish non-complying work that demonstrated high effort} 

Assignment to $OK$ had no effect on subjects' accept/reject
recommendations: 
\begin{align} \label{eq:E.approval}
  approve = \underbrace{-0.002}_{[0.041]}\cdot OK + 
\underbrace{0.872}_{[0.028]}\end{align}
with $n=272$ and $R^2 = 0$. For transfers
in the dictator game: 
\begin{align}
  bonus =
  \underbrace{-0.397}_{[0.263]}\cdot OK +
  \underbrace{5.305}_{[0.182]}\end{align} with
$n=272$ and $R^2 = 0.01$.  Note that the bonus
level itself was quite high, and most subjects transferred a more than
equitable split. Given the strong positive correlation between
subjects' own productivity and self-serving behavior in the dictator
game, the large average bonus size is consistent with the fact that
most subjects showed low productivity on the initial task (because
they were induced to choose $y_1=2$). Although the coefficient on $OK$
in the regression above is not significant, it is not a precisely
estimated zero, as in the case of approval in Equation
\ref{eq:E.approval}. Self-serving behavior among subjects assigned to
$OK$ is concentrated among highly productive types evaluating the
$2$-label evaluated work. We can see this by the large negative
coefficient on the group/productivity ($y_1 \times OK$) interaction
term:

\begin{align} bonus = \underbrace{0.720}_{[0.502]}\cdot OK + 
 \underbrace{0.020}_{[0.152]}\cdot y_1 + 
 \underbrace{-0.547}_{[0.204]} \cdot y_1 \times OK + 
\underbrace{5.264}_{[0.383]}\end{align}
with $n=272$ and $R^2 = 0.05$.

\subsubsection{Exposure to highly productive work increased output even when expectations were explicit} 
Workers exposed to $OVER$ increased their output compared to those
exposed to $OK$:
\begin{align} y_2 = \underbrace{0.658}_{[0.120]} \cdot y_1 + 
 \underbrace{-0.659}_{[0.183]} \cdot OK + 
  \underbrace{1.294}_{[0.299]}  
\end{align} 
with $n = 272$ and $R^2 = 0.25$. However, unlike
in Experiment D, this effect was not conditional upon initial
output. When the regression above is augmented with an $y_1 \cdot OK$
interaction term (not shown), the coefficient on the interaction is
small and insignificant, which is consistent with there being little
variation in initial output (i.e., the distribution of $y_1$ is
heavily concentrated at $y_1=2$).

\subsection{Discussion} 
It is clear that many workers take explicit requests from employers as
informative and worth complying with. Yet a substantial number of
workers remained susceptible to peer effects that could, in principle,
have led them to have their work rejected for not following the letter
of the instructions. There are several possible explanations.

One possibility is mistake. Workers might believe that they
misinterpreted the instructions or that other workers have some inside
knowledge. Workers might reasonably believe that we had free disposal
of extra labels and added a few more to provide a margin of safety.
However, the requirement was clearly stated at least three times to
workers, and many initially complying workers were still pulled upward
by highly productive peers.

Another possibility is that workers want to produce an amount
comparable to that of their peers, regardless of employer
instructions. Given the propensity of peers to punish low effort,
matching the output of one's peers is a good idea if there is a chance
one will be evaluated by those peers. Because subjects are asked to
evaluate workers, it seems likely that they infer they will also be
evaluated by other workers. Given that other workers are likely to
reward or punish based on apparent effort, not necessarily on
compliance with the stated standards, above-standard output could be
rational even if it is technically non-complying.

\section{Conclusion} 
This paper reports a number of results on punishment, productivity and
peer effects: (1) workers readily punish low-effort work; (2) workers
are susceptible to peer effects, but the effects are conditional upon
a worker's productivity; (3) a worker's willing to punish is mediated
by their own productivity, which is in turn malleable; (4) workers
punish low effort, not failure to comply with an employer's
instructions. Some of these findings contradict or at least complicate
results from other workplace settings. Future research should
investigate the generalizability of these findings and which should
clarify which findings are general features of how humans think about
work and which are context-specific.

If strong reciprocity is a general feature of human organizations,
then a natural question for managers is whether they should encourage
this phenomenon among workers. Here, context seems to matter
greatly. Giving workers thicker sticks or juicier carrots to use on
their peers may backfire if workers are enforcing norms contrary to
the best interests of the firm. Further, other research has shown that
workers will enforce norms that are directly counter productive (e.g.,
Roy's \citeyear{roy1952quota} work on machine shops). Creating the
tools for norm enforcement is risky when knowledge of what will be
enforced is murky.

A key theme of these experiments is the apparent pliability of
productivity, which highlights the danger of what might be called
``organizational alchemy,'' i.e., the attempt to harness the power of
peer effects without knowing how they work in the relevant
context. For example, in contrast to several other findings, ``bad''
workers did not improve after evaluating good work.  At least in the
context examined, it would be a mistake to mix workers of different
abilities with the hope that the good would raise the bad.

\bibliographystyle{aer}
\bibliography{peer_effects.bib}

\end{document}